# BEHAVIOR OF PLASMA AND FIELD PARAMETERS AND THEIR RELATIONSHIP WITH GEOMAGNETIC INDICES DURING INTENSE GEOMAGNETIC STORMS OF SOLAR CYCLE 23

Navin Chandra Joshi, Neeraj Singh Bankoti, Seema Pande[a], Bimal Pande*, Kavita Pandey

Department of Physics, D.S.B. Campus, Kumaun University, Naini Tal – 263 002, Uttarakhand, India

[a]Department of Physics, M.B.P.G. College, Haldwani, Kumaun University, Naini Tal, Uttarakhand, India

E-mail address: njoshi98@gmail.com; nsbankoti@gmail.com; pande.seema@gmail.com; kavitaphysics@yahoo.co.in

*Email: pandebimal@yahoo.co.in; Tel.: +91 9412044061; Fax: +91 5942237450.

## ABSTRACT

A correlative study between the geomagnetic indices and the peak values of various plasma and field parameters during rising, maximum and decay phases as well as during complete solar cycle 23 have been presented. We have also presented the lag/lead analysis between the maximum of Dst and peak values of plasma and field parameters and found that peak values of lag/lead time lies in the ±10 hr interval. Three geomagnetic storms (GMSs) and associated solar sources observed during these phases of this solar cycle have also been studied and found that GMSs are associated with large flares and halo CMEs.

# 1. INTRODUCTION

When an intense and long lasting interplanatry convective electric field leads through substantial energization in the magnetosphere-ionosphere system to an intense ring current which is stronger than the threshold of the quantifying storm time Dst (disturbance storm time) index, the time interval is defined as a geomagnetic storm (GMS). GMSs are usually classified by the Dst indices as intense storms (peak Dst ≤ −100 nT), moderate storms (−100 nT < peak Dst < −50 nT) and weak storms (peak Dst > −50 nT) (Gonzalez et al., 1994). In terms of time sequence, a GMS can be described in three phases: the initial, the main and the recovery phase. The initial phase may be gradual, or be represented by an abrupt change in the Dst, called a sudden commencement. The main phase of a storm is said to begin when the Dst assumes negative values and ends when it reaches its minimum decrease. The recovery phase, usually the longest one, is characterized by the returning of Dst to its pre-sudden commencement values. During a GMS, the Sun and the magnetosphere are connected giving rise to several changes both in interplanetary space and terrestrial environment. All the perturbations during GMSs involve energy transfer from solar wind to magnetosphere-ionosphere system and modify plasma and magnetic field there. Efficiency of a process seems to depend on the southern component of magnetic field and on the solar wind speed i.e., on the dawn-dusk component of solar wind electric field (Gonzalez et al., 1994).

Dst, Kp, ap and AE indices are the four most commonly used geomagnetic indices (GI). Dst index is defined as the hourly average of the deviation of H (horizontal) component of magnetic field measured by several ground stations in mid to low latitudes and represents the degree of equatorial magnetic field deviation specifying the magnitude of GMSs. This is measured in the units of nano tesla (nT). Kp index represents the intensity of planetary magnetic activity as seen at subauroral latitudes and is given for every 3-h interval. The K index for each of the contributing mid-latitude observing stations reflects the maximum range of any component of the field over the 3-h interval at each station. The Kp index is the average of the K values from all contributing observatories. A conversion scale transforms the quasi-logarithmic Kp to a linear index named ap index. The state of magnetosphere is described by different indices and Dst and Kp indices are usually used for identification of magnetic storms (Mayaud, 1980). AE is defined by Davis and Sugiura (1966), to measure primarily the variations in the auroral electrojets. It is based on 1-min values of the H component trace from auroral-zone observatories located around the world. In general, the AE index is used to describe the substorm intensity. Substorms, as one of the most important processes in the Earth's magnetosphere, receive wide attention in the space community because of their large scale influence on other magnetospheric and ionospheric dynamic processes in the global environment of the Earth (Davis and Sugiura, 1966).

Different types of studies of GMSs and disturbances in the geomagnetic activity have been carried out in the past to understand the solar-terrestrial relationships and to ascertain those factors that are ultimately responsible for GMSs (Burlaga et al., 1982; Gonzalez and Tsurutani, 1987; Gonzalez et al., 1990; Gonzalez et al., 1994; Saba et al., 1997; Tsurutani and Gonzalez, 1997; Badruddin, 1998; Gonzalez et al., 1999; Gosling and Pizzo, 1999; Crooker, 2000; Papitashvili et al., 2000; Srivastava and Venkatakrishnan, 2002; Wang et al., 2002; Echer et al., 2005; Yermolaev et al., 2005; Gonzalez et al., 2007; Kane and Echer, 2007; Zhang et al., 2007; Echer et al., 2008; Gopalswamy, 2008; Khabarova and Yermolaev, 2008 and reference therein).

The study of GMSs is one of the main areas of space weather research. A brief review of magnetospheric and interplanetary phenomena at intervals with enhanced solar wind-magnetosphere interaction has been presented by Gonzalez et al., (1994). It is assumed that Sun-Earth interaction depends on solar wind. In fact intense GMSs seem to be related to intense interplanetary magnetic field (IMF) and its southern component for a longer time (Gonzalez and Tsurutani, 1987). Gonzalez et al., (1999) presented a brief study about the interplanatry origin of GMSs and found that two interplanatry structures are important for the development of storms; the sheath region just behind the forward shock, and coronal mass ejection (CME) ejacta involving intense southward IMFs.

In the study of solar-terrestrial relationships, geomagnetic activity indices play an important role. Three-hourly average values of the Dst, AE and ap geomagnetic activity indices have been studied by Saba et al., (1997) for 1 year duration each near the solar minimum (1974) and at the solar maximum (1979). They found out that in 1979 seven intense GMSs (Dst < -100 nT) occurred, whereas in 1974 only three were reported and also that the yearly average of AE is greater in 1974 than in 1979, the reverse seems to be true for the yearly average of Dst. Papitashvili et al., (2000) studied solar cycle effects in planetary geomagnetic activity in which 27-day averages of several plasma and field parameters (from OMNI data base) are compared with equivalent Kp and Dst averages and concluded that changes in the magnitude rather than in direction are the cause of primary solar cycle variations in the IMF.

Interplanetary phenomena have been classified into six categories; 1) heliospheric current sheet 2) slow solar wind from coronal streamers 3) fast solar wind from coronal holes 4) compressed streams of solar wind (corotating interaction region (CIR) and streams ahead of magnetic clouds) 5) magnetic clouds (ejecta) and 6) decompressed streams of solar wind. Of these only CIR and magnetic cloud are geoeffective because they may include long southward Bz component of IMF (Gosling and Pizzo, 1999; Gonzalez et al., 1999; Crooker, 2000; Yermolaev and Yermolaev, 2006). Echer et al., (2008) investigated that during the rising phase of the solar cycle magnetic clouds which drove fast shocks and sheath fields are the dominant structures causing intense storms. At solar maximum, sheath fields, followed by combined sheath and magnetic clouds and then by magnetic clouds which drove fast shocks were responsible for most of the storms. During the declining phase magnetic clouds which drove fast shocks, sheath fields and CIRs are the main interplanetary structures leading to intense storms. Khabarova and Yermolaev (2008) found out that the solar wind behavior before and after the onset of all magnetic storms is different form the well-known behavior of the solar wind before and after severe magnetic storms and the well-known rule of high-speed stream geoeffectiveness does not work for most GMSs.

Interplanatry coronal mass ejection (ICME) is the general name given to various types of Interplanatry structures resulting from CMEs. Magnetic clouds (MCs) are a subset of ICMEs when they have enhanced magnetic field, smooth magnetic field rotation, and low plasma beta as defined by Burlaga et al., (1982). Echer et al., (2005) presented a statistical study of MC parameters and geoeffectiveness which is based on the analysis of 149 magnetic clouds during the period 1966–2001 and found that overall 77% of MCs are geoeffective in the sense that they are followed by intense or moderate magnetic storms with the percentage of peak southward magnetic field within the clouds reaching 70% of the total magnetic field. Gopalswamy (2008) has also presented an extensive discussion on the geoeffectiveness of MC sheaths.

A number of studies to establish correlations between the GI and the various plasma and field parameters and for developing models for predicting the occurrence of GMSs which are important for space weather predictions have been performed in the past (Snyder et al., 1963; Crooker and Gringauz, 1993; Wu and Lundstedt, 1997; Gonzalez et al., 1998; Ondoh, 2000; Jurac et al., 2002; Wu and Lepping, 2002; Srivastava and Venkatakrishnan, 2004; Gonzalez and Echer, 2005; Kane, 2005; Srivastava, 2005a, 2005b; Singh et.al., 2006; Kane and Echer, 2007; Echer et al., 2008; Mansilla, 2008). Snyder et al., (1963) studied a possible link between Kp index and solar wind speed V and stated that the relationship of V with Kp was not precise but only suggestive. Crooker and Gringauz (1993), performed a correlative study between the GI (Ap, Dst indices) and plasma parameters (solar wind speed, IMF) and products of plasma and field parameters during solar cycles 20 and 21. Gonzalez and Echer, (2005) studied the relationship between the Dst and Bz component of the IMF for 64 intense GMSs (Dst ≤ - 85 nT) during the period 1997–2002 and found that the Bz value at peak Dst is ≈75% of the peak Bz value in the entire event. Kane, (2005) studied the relationships between Dst, solar wind speed V and Dst, VBz (product of V and southward component Bz of IMF) for several events during 1973–2003, particularly to check space weather models based on Dst-V relationship and found that, VBz and not V is the appropriate variable relevant for Dst change. They also investigated that moderate or strong GMSs occurred only when V was above ≈ 350 km/s. Srivastava, (2005a, 2005b) developed a logistic regression model which can be implemented for predicting intense/super-intense GMSs. Recently Mansilla (2008), presented a correlative study between the intensity of a GMS and the peak value reached by some solar wind parameters (solar wind speed V and solar wind number density D) and the southward component of the IMF from 2000 to 2005 and found that peak Dst is correlated to the maximum negative component Bz of the IMF better than the maxima of solar D and V respectively.

Srivastava and Venkatakrishnan, (2004) examined the solar origins of the geoeffective CMEs and their interplanetary effects, namely, solar wind speed, interplanetary shocks, and the southward component of the IMF in order to investigate the relationship between solar and interplanetary parameters. Their results also show that the intensity of GMSs depends most strongly on the southward component of the IMF, followed by the initial speed of the CME and the ram pressure in this order. Gupta and Badruddin (2008) using superposed epoch analysis presented a relative contribution of various types of structures, the average behavior of various plasma and field parameters during, before and after the passage of five types of interplanetary structures responsible for major storms. They also presented a lag/lead time analysis of various plasma and field parameters. In order to understand the response of the magnetosphere to interplanetary conditions during GMSs, several studies have derived relations between interplanetary values of various plasma and field parameters and various combinations of these parameters. A unique relationship which may ultimately lead to an unambiguous understanding of the phenomenon and predict the occurrence of GMSs is yet to emerge.

Study of identification of solar sources (i.e. active region, flare and CME) associated with GMSs has been performed by various authors in recent years to find out the connection between the Sun and the Earth's atmosphere (Tsurutani and Gonzalez, 1997; Wang, et al., 2002; Zhang et al., 2003; Dal Lago, et al., 2004; Srivastava and Venkatakrishnan, 2004; Correia and de Souza, 2005; Wang, 2007; Zhang et al., 2007; Echer et al., 2008). The geospheric environment is highly affected by the Sun and its features such as solar flares, active prominences, disappearing filaments, CMEs etc. which are responsible for some large/small GMSs (Gonzalez et al., 1994). CMEs and CIRs are the two large scale interplanetary structures that cause GMSs (Wang, 2007). CMEs and high-speed solar wind streams (HSS) are two solar phenomena that produce large scale structures in the interplanetary medium. CMEs evolve into ICMEs and the high speed stream (HSS) results in CIR when they interact with preceding low solar wind. Detailed studies of solar cycle 23 have enriched data sets that reveal various aspects of GMSs in an exhaustive detail both at the Sun where the storm causing disturbances originate and in the geospace where the effects of the storms are directly felt. Zhang et al., (2003) identified the solar CME sources for 27 major GMSs (Dst≤ -100 nT) which occurred between 1996 and 2000 and found that most geoeffective CMEs originate within a latitude strip of ± 30°. They also concluded that whether these geoeffective CMEs are either full-halo CMEs (67%) or partial-halo CMEs (30%), there is no preference for them to be fast CMEs or to be associated with major flares and erupting filaments. Correia and de Souza (2005) identified solar CME sources for selected major GMSs (Dst ≤ -100 nT) that occurred in October–November 2003 and reported that the fast and large CMEs propagating in a disturbed solar wind could accelerate energetic particles and intensify the magnetic storms. Zhang et al., (2007) presented the results of an investigation of the sequence of events from the Sun to the Earth that ultimately led to 88 major GMSs (Dst ≤ -100 nT) that occurred during 1996–2005. They also identified and characterized each major GMS, the overall interplanatry source type, time, velocity, and angular width of the source CME, type and heliographic location of solar source region, the structure of the transient solar wind flow with the storm-driving component specified, arrival time of shock/disturbance, and the start and end times of the corresponding ICMEs.

Cliver et al., (2009) report on the solar source of the great GMS (Dst = -354 nT) on 8-10 November 1991. The solar source is identified as the large-scale eruption of a long solar filament followed by a soft X-ray arcade, which was found to rank 15th on a list of Dst storms from 1905 to 2004. Jordanova et al., (2008) studied the effect of electromagnetic ion cyclotron wave scattering on radiation belt electrons during the large GMS of 21 October 2001 (Dst = -187 nT) using their global physics-based model.

The study of magnetic storms is one of the main ingredients of space weather. These storms generate several changes both in interplanetary space and terrestrial environment and can damage the power supply, radio communications and spacecrafts. In the present communication, results are presented for the relationships between GI and several plasma and field parameters for a 91 GMSs selection of solar cycle 23. In Section 2 data sources and selection criteria of data set have been presented. In Section 3 analyses and results with correlation between GI and plasma and field parameters (Section 3.1) and study of three GMSs from the rising, the maximum, and the decay phases and associated solar sources have been investigated (Section 3.2). Discussion of results and final conclusions are presented in the last Section 4.

## 2. DATA SET AND STATISTICAL TOOLS

Conditions in the solar wind resulting in magnetic storms on the Earth have been a subject of long and intensive investigation. The solar wind plasma and field measurements with 1 h time resolution were obtained from the OMNI website as: http://omniweb.gsfc.nasa.gov. Hourly Dst indices were obtained from the world data center at the University of Kyoto database as: http://swdc.kugi.kyoto-u.ac.jp/dstdir. OMNI data center provide field magnitude average (FMA) ($|B|$ (nT)), magnitude of average field vector also known as total magnetic field (Bt (nT)) of IMF, negative y-component of IMF (By (nT)), negative z-component of IMF (Bz (nT)), variance of total IMF ($\sigma_B$ (nT)). All the field parameters are in geocentric solar ecliptic (GSE) coordinate system. They also provide proton temperature (T ($^oK$)), proton density (D ($cm^{-3}$)), plasma speed (V (km/s)), flow presser (P (nPa)), electric field (E (mV/m)), plasma beta (β), Kp index, Dst index, AE index and ap index. We have selected 91 GMSs occurring during solar cycle 23. Out of these 90 are the same events as chosen by Zhang et al., (2007) and Echer et al., (2008). We define a major GMS as a minimum in the hourly Dst index falling below ≤ -100 nT. A similar threshold for major/intense storms has been reported by other authors (Tsurutani and Gonzalez, 1997; Zhang et al., 2007). For all 91 events we have found out the peak values of GI and various plasma and field parameters. We have performed a linear regression analysis of the form Y = A + BX where Y is the peak value of GI and X is the peak value of various plasma and field parameters. In the above data base of 91 GMSs there are four events for which AE index is not given, hence for the AE index analysis we have used only 88 GMSs. In order to find out possible relationships between peak values of GI and various plasma and field parameters in different phases we have applied the above regression analysis in different phases of solar cycle 23. For this we have divided the complete cycle into three phases; the rising (1996-1999), the maximum (2000-2002) and the decay (2003-2008) phase.

In the present work we have also made an attempt to find solar sources of the three intense GMSs. For this we have used observations by the Large Angle Spectroscopic Coronagraph (LASCO), Michelson Doppler Imager (MDI) and by the Extreme Ultraviolet Imaging Telescope (EIT) onboard Solar and Heliographic Observatory (SOHO) competent to image different solar surfaces at different wavelengths. Solar source data have been downloaded from various catalogs available on internet.

Sunspot data from SOHO/MDI (http://soi.stanford.edu/data/),

flare data from SOHO/EIT (http://umbra.nascom.nasa.gov/eit/eit-catalog.html), and

CME data from SOHO/LASCO (http://cdaw.gsfc.nasa.gov/CME_list/).

We have calculated the $10^{th}$ percentile ($P_{10}$), which state that 90% of the events have values larger than $P_{10}$.

Average, median and standard deviation of peak values of various GI, plasma and field parameters as well as average, median, mode and standard deviation of lag/lead time of these parameters with respect to GMSs peak time during the rising, the maximum and the decay phase and total solar cycle 23 has also been presented.

## 3. ANALYSES AND RESULTS

### 3.1. CORELATION BETWEEN GEOMAGNETIC INDICES AND VARIOUS PLASMA AND FIELD PARAMETERS

A table of selected 91 intense GMSs observed during solar cycle 23, with their associated interplanetary (field/plasma) parameters is given in Table 1. First column represents the number of events. Date of Dst having its maximum value and the corresponding time is represented in second and third columns respectively. From fourth to fourteenth columns peak values of various plasma and field parameters ($|B|$, Bt, By, Bz, $\sigma_B$, T, D, V, P, E and β) have been presented. Next four columns show peak values of four GI (Kp, Dst, AE and ap). Last column represents the Solar wind-magnetospheric coupling parameter (VBz). Missing Peak values are marked as ****.

Yearly variation of CMEs, Hα solar flares, soft X-ray (SXR) solar flares, sunspot numbers (SNs) and GMSs during solar cycle 23 is represented in Fig. 1(a). It is clear from this figure that the variation of GMSs is similar to the CMEs, Hα flares, SXR flares and SNs. There is a significant peak in all the rows observed during maximum phase (2000-2002). Maximum number of GMSs has been observed during maximum phase of solar cycle 23 in 2001 and 2002 (14 GMSs). During the rising phase maximum number of GMSs observed in 1998 (12 GMSs) and during the decline phase in 2005 (10 GMSs). It is also clear that after 2006 no intense GMSs have been observed. Fig. 1(b) shows pi chart representing the distribution of intense GMSs during the rising, maximum and the decay phase of solar cycle 23. It shows that the maximum number of GMSs occurred in the maximum phase (43.96%) whereas in the rising phase the number of intense GMSs is minimum (25.27%). To understand the relation between the intense GMSs with solar activity features we have used linear regression analysis using yearly values shown in Fig. 2. A good correlation between the yearly values of GMSs and SNs (correlation coefficients R = 0.777) (Fig. 2a), SXR (R = 0.877) (Fig. 2b) and Hα solar flares (R = 0.663) (Fig. 2c) have been obtained. However CMEs show a moderate correlation (R = 0.513) with GMSs (Fig. 2d). From the Fig. 1(a) and Fig. 2 we have concluded that intense GMSs have a good correlation with solar activity features.

Relationship among different GI for 91 intense GMSs during 1996–2008 is presented in Fig. 3. The correlation coefficients between peak Dst-Kp, Dst-ap and Dst-AE index come out to be -0.716, -0.820, -0.554 respectively (Fig. 3a, 3b and 3c). Correlation coefficients between peak AE-Kp, AE-ap index is found to be 0.731 and 0.764 respectively (Fig 3d and 3e). From these results it is clear that Peak Dst is well correlated (negative) with ap and Kp indices and moderately correlated (negative) with AE-index whereas AE-index is well correlated (positive) with Kp and ap indices.

Figs. 4, 5, 6 and 7 represent the correlation between peak values of Dst, ap, Kp and AE indices and all plasma and field parameters respectively. The linear regression equations and the corresponding correlation coefficients are given in figures. It can be seen form Fig. 4 that Peak Dst index is well correlated with $|B|$ (R = -0.772), Bt (R = -0.797), Bz (R = 0.785), E (R = -0.721) and with VBz (R= 0.817) (Figs. 4a, 4b, 4d, 4j and 4l) whereas it is moderately correlated with By (R= 0.505), $\sigma_B$ (R = -0.605) and V (R = -0.531) (Fig. 4c, 4e and 4h). The correlation between peak Dst index with T (R = -0.379), D (R = -0.031), P (R = -0.376) and β (R = 0.026) (Figs. 4f, 4g, 4i and 4k) is poor. It is clear form this figure that Dst index is well correlated with $|B|$, Bt, Bz, E and VBz. Peak value of ap index shows a good correlation with $|B|$ (R = 0.770), Bt (R = 0.766), Bz (R = -0.663), $\sigma_B$ (R =0.701), V (R = 0.741), E (R = 0.714) and VBz (R= -0.782) (Figs. 5a, 5b, 5d, 5e, 5h, 5j and 5l) and shows a weak correlation with D (R = 0.009) and β (R = -0.047) (Figs. 5g and 5k). ap index shows moderate correlation with By (R = -0.478), T (R= 0.610) and P (R= 0.517) (Figs. 5c, 5f and 5i). It is also clear that ap index is well correlated with $|B|$, Bt, Bz, $\sigma_B$, V, E and VBz. It can be seen from Fig. 6 that the Kp index has a good correlation with $|B|$ (R= 0.727), Bt (R= 0.717), $\sigma_B$ (R= 0.663), V (R= 0.701) and VBz (R= -0.682) (Figs. 6a, 6b, 6e, 6h and 6l) whereas it is weakly correlated with peak values D (R = 0.059) and β (R = -0.012) (Figs. 6g and 6k). Kp index shows moderate correlation with By (R = -0.457), Bz (R= -0.592), T (R= 0.580) and P (R= 0.539) and E (R= 0.648) (Fig. 6c, 6d, 6f, 6i and 6j). Thus it is evident that Kp index shows good correlation with $|B|$, Bt, $\sigma_B$, V and VBz. AE index does not show good correlation with plasma and field parameters. AE index shows good correlation with V (R = 0.694) (Fig. 7h) whereas it shows moderate correlation with FMA (R = 0.647), Bt (R = 0.623), Bz (R= -0.491), $\sigma_B$ (R= 0.606) T (R= 0.560), P (R= 0.515) E (R= 0.511) and VBz (R= -0.610) (Fig. 7a, 7b, 7d, 7e, 7f, 7i, 7j and 7l) whereas a poor correlation is observed between AE index with By (R= -0.379), D (R= -0.067), β (R= -0.051) (Fig. 7c, 7g and 7k).

Table 2 shows the correlation coefficients of four GI with 11 plasma and field parameters and VBz during three different the rising, the maximum and the decay phase of solar cycle 23. It is observed that the correlation coefficients between the GI and various plasma and field parameters are different in all the three phases. Values of correlation coefficients between Dst index with $|B|$, Bt, By, Bz and VBz is maximum in decay phase while for $\sigma_B$ and D it is maximum in maximum phase and for T, V, P, E, β it is more in the rising phase. Values of correlation coefficients between Kp index and $|B|$, Bt, By, Bz, $\sigma_B$, and D are maximum in maximum phase. Values of correlation coefficients for Kp index with VBz and T are maximum in decay phase and with V, P, E, β are maximum in the rising phase. Values of correlation coefficient between ap index with $|B|$, Bt, Bz, $\sigma_B$, D and VBz are maximum during maximum phase while for By it is maximum in decay phase and for T, V, P, E, β these are large in the rising phase. Finally the value of correlation coefficients between AE index and $|B|$, Bt, By, Bz, T, P and VBz are maximum in maximum phase. Value of correlation coefficient for $\sigma_B$ is maximum in the decay phase and for D, V, E, β it is large in the rising phase.

Distribution of peak value of GI and various plasma and field parameters have been represented in Figs. 8 and 9 respectively. The average and median values of the GI and various plasma and field parameters during the considered time period are marked on the respective figures with solid and broken lines respectively.

In Table 3 we have presented a comparison of average and median values of four GI and the various plasma and field parameters. $P_{10}$ values of all the plasma, field parameters and GI for the total cycle have also been represented in this table. From this Table it is clear that the maximum average and median of peak values of four GI lie in the decay phase. The maximum average and median value of peak values of $|B|$, Bt, By, Bz, T, V and E are maximum in decay phase of solar cycle under investigation whereas $\sigma_B$, P and β are maximum in the maximum phase. D has its maximum, average and median of peak values in the rising phase only. Since the average and median values of GI and various plasma and field parameters are large in decay phase hence we may conclude that the decay phase of solar cycle 23 opens up new vistas with regard to the behavior of various plasma and field parameters as well as origin of GMSs.

Average, median and mode values of Lag (-)/lead (+) time (hr) of maximum in various GI and plasma and field parameters with respect to GMSs peak time (i.e. minimum Dst) are represented in Table 4. Negative (-) sign represents the peak value obtained prior to minimum Dst while positive (+) sign shows the peak value obtained after the minimum Dst. Average values of time difference between minimum Dst and peak values of Kp, ap and AE indices come out to be negative for all the phases of solar cycle 23. During Maximum phase these average values come out to be higher (more negative) than the other phases. Average and median values of $|B|$, Bt, Bz, $\sigma_B$, P and E come out to be negative during all the three phases of solar cycle 23 whereas for T, V and β these values come out to be positive for all the three phases. Time difference value of D with minimum Dst shows positive values during rising and negative values in maximum and decay phase. By shows positive values for rising and maximum phase and negative values in decay phase. Overall, the peak values of Kp, ap, AE, $|B|$, Bt, Bz, $\sigma_B$, D and E commence before the peak Dst value and hence these parameters may play a major role in predicting the intense GMSs.

Distribution of Lag /Lead times of different GI and various plasma and field parameters with respect to peak GMSs (i.e. minimum Dst) have been presented in Figs. 10 and 11 respectively. A peak at zero lag and lead time is obtained for ap and Kp index (Fig. 10a). Peak value of AE-index for most of the intense GMSs obtained around 0-10 hr prior to the peak Dst is observed (Fig. 10b). From Figs.10 and 11 it can be seem that for both GI and various plasma and field parameters there is a peak in lag/lead time from -10 to +10 (hr).

### 3.2. STORMS ON RISING, MAXIMUM AND DECAY PHASES OF SOLAR CYCLE 23

#### 3.2.1. NOVEMBER 7, 1997 GMS

In Fig. 12, in the left panel is a composition of solar, interplanetary and geomagnetic observation of intense GMS on November 7, 1997 (event number 5 in Table 1), showing images from MDI, EIT and LASCO. In the right panel of this figure from top to bottom we have plotted the IMF parameters ($\sigma_B$, Bz, By, Bt and $|B|$), GI (ap, AE, Dst, and Kp indices) and plasma parameters (β, E, P, V, D and T) for a period of 5 days from November 6 to November 10, 1997 (all the parameters are in GSE coordinates). Active region, date, start time, importance, brightness class, intensity class, latitude and longitude of the associated flare are represented in Table 5. The date, start time, width, and speed of associated CME are also represented. Peak value of Dst index has also been mentioned in the same table. Active region NOAA AR 8100 produced X2.1, 2B Class flares on 04/11/1997, 05:52 UT. This flare produced a halo CME on 04/11/1997 at 06:10 UT with a linear speed of 785 km/s which gave rise to interplanetary shocks recorded by CELIAS on board SOHO (http:umtof.umd.edu/pm) and produced an intense GMS (Dst = -110 nT). The right panel of Fig. 12 clearly shows the solar wind features of a single ICME, which is composed of a shock and a magnetic cloud. Sudden commencement of the storm (SSC) occurred at 22:48 UT on November 6 (http://www.ngdc.noaa.gov/stp/SOLAR/ftpSSC.html). At this moment, all major solar wind parameters including velocity, density, and magnetic field showed a jump up. Following the SSC, the Bz component of the IMF oscillated rapidly for several hours, then turned north for several hours and finally turned to south around 07:00 UT on November 7, 1997. The Dst index rapidly decreased following the south turning of Bz component of IMF and reached a minimum value of -110 nT at 5:00 UT on November 7, 1997.

### 3.2.2. OCTOBER 21, 2001 GMS

Fig. 13 is a composition of solar, interplanetary and geomagnetic observations of intense GMS on October 21, 2001 (event number 45 in Table 1) showing images from MDI, EIT and LASCO instruments in the left panel and the behavior of various GI and various Plasma and field parameters in the right panel. Details about the associated active region, flare and CME are described in Table 5. Date, start time, importance class, brightness class, intensity class, latitude and longitude of the associated flare are represented in the same table. The relevant observations of associated CME are also represented. Active region NOAA AR 9661 produced large X1.6, 2B Class flare. The region was well-positioned to produce Earth directed CME events. This flare (X1.6 2B) is associated with a CME on 19/10/2001 at 16:50 UT and CME gave rise to interplanetary shocks which were recorded by CELIAS on board SOHO (http:umtof.umd.edu/pm) and produced an intense GMS (Dst = -187 nT). The SSC occurred at 16:48 UT on October 21 (http://www.ngdc.noaa.gov/stp/SOLAR/ftpSSC.html). Following the initial shock the solar wind speed gradually decayed. The Bz component of the IMF oscillated rapidly for several hours after this SSC then turned north for several hours, and finally turned to south around 22:00 UT on the same day. The Dst index rapidly decreased following the south turning of Bz component of IMF and reached a minimum value of -187 nT at 22:00 UT on October 21, 2001.

### 3.2.3. DECEMBER 15, 2006 GMS

Fig. 14 is a composition of solar, interplanetary and geomagnetic observation of intense GMS on December 15, 2006 (event number 91 in Table 1), showing images from MDI, EIT and LASCO instruments in the left panel and the behavior of various plasma and field parameters in the right panel. Intense GMS on December 15, 2006 associated with active region, flare and CME are described in Table 5. Date, start time, importance class, brightness class, intensity class, latitude and longitude of the associated flare are represented in the same table. The date, start time, width, and speed of associated CME are also represented. The region was well-positioned to produce Earth directed CME event. An X3.4 4B flare in the early hours of December 13, 2006 (02:14 UT) produced a CME which was Earth-directed and capable of producing large GMS activity starting in the late hours of Dec 14 and early Dec 15. This CME occurred on December 13, 2006 (02:54 UT) giving rise to interplanetary shocks recorded by CELIAS on board SOHO (http:umtof.umd.edu/pm) and produced an intense GMS (-146 nT). A sudden increase in the solar wind speed occurred around 14:00 UT on December 14 indicating the arrival of a shock. The SSC occurred at 14:14 UT on 14 December (http://www.ngdc.noaa.gov/stp/SOLAR/ftpSSC.html). Following the initial shock the solar wind speed gradually decayed. Following the SSC, the Bz component of the IMF oscillated rapidly for several hours then turned north for several hours and finally turned to south around 23:00 UT on December 14. The Dst index rapidly decreased following the south turning of Bz component of IMF and reached a minimum value of -147 nT at 8:00 UT on 15 December.

## 4. DISCUSSIONS AND CONCLUSIONS

In the present work an effort has been made to achieve a better understanding of the geomagnetic indices and their relationships with various plasma and field parameters during three different phases (rise, maximum and decay) as well as total solar cycle 23 by means of their averages and correlations. The main conclusions can be summarized as following.

- From Fig. 1(a) and Fig. 2 it is clear that there is a good correlation between intense GMSs and solar activity features (i.e. sunspot numbers, soft X-ray flares, Hα solar flares, CMEs).
- From Fig. 3 it can be seen that the peak values of all four GI are highly correlated with each other with correlation coefficient $R \geq 0.7$ except Dst vs AE ($R = -0.554$).
- Peak values of Dst index are well correlated ($R \geq 0.7$) with peak values of $|B|$, Bt, Bz, E and VBz (Fig. 4).
- Peak values of ap index are well correlated with peak values of $|B|$, Bt, Bz, $\sigma_B$, V, E and VBz (Fig. 5).
- Peak values of Kp index are well correlated with peak values of $|B|$, Bt, $\sigma_B$, V and VBz (Fig. 6).
- Peak value of AE index is well correlated with peak values of V (Fig. 7). Finally from the above four observations it can be concluded that the peak values of GI are in good correlation with $|B|$, Bt, Bz, $\sigma_B$, V, E and VBz and hence these parameters are most useful for predicting GMSs and substorms during the rising, the maximum, the decay phases and also the total solar cycle 23.

- Comparative analyses of correlation coefficients of Dst with field parameters show that during decay phase Dst index is more correlated with field parameters whereas it is well correlated with plasma parameters during rising phase. Kp, ap, and AE indices show maximum correlation with field parameters during maximum phase while these GI show maximum correlation with plasma parameters mostly in the rising phase (Table 2).
- Average and peak values of GI as well as many plasma and field parameters show maximum values in decay phase of solar cycle 23 (Table 3). Hence study of intense GMSs and behavior of plasma and field parameters during the decay phase of solar cycle 23 yields new information about the origin of GMSs and their effect on the Earth's atmosphere.
- Lag/lead time analyses of different GI and various plasma and field parameters with minimum Dst show that the peak values of lag/lead time lie in the ±10h (Figs. 10 and 11).
- All three GMSs (November 7, 1999; October 21, 2001 and December 6, 2006) are associated with large flares (≥ X class) and halo CMEs.

It is important to identify the solar drivers of the geomagnetic activity in order to be able to predict the occurrence of a strong GMS. Several workers have discussed this aspect in great detail over various periods of the solar activity cycle (Gosling et al., 1991; Tsurutani and Gonzalez, 1997; Richardson et al., 2000; Richardson et al., 2001). On the basis of these studies, one expects a large number of intense GMSs (- 200 nT < Dst < - 100 nT) close to the solar maximum than during the minimum. Our results confirm this inference for solar cycle 23 as maximum phase of cycle 23 produced a large number of GMSs (43.95%) compared to the rising (25.27%) and the decay (30.77%) phases. We also obtained the approximate rate of intense magnetic storms for different phases of solar cycle 23; rising phase ≈ 5.75 storm/year, maximum phase ≈ 13.33 storm/year and decay phase ≈ 4.66 storm/year. Last row of Fig. 1(a) represents the solar cycle distribution of intense storms. The peaks at 2001-2002 and 2004-2005 in intense storms correspond to the solar cycle maximum and the declining phase respectively (Fig. 1(a)). Occurrence of peaks during maximum and decay phase has been reported for several other solar cycles also by Gonzales et al., (1990).

Wu and Lundstedt (1997) and Wu and Lepping (2002) studied the correlations of Dst with the maximum of solar wind speed V and Bz component of IMF and found that Bz component is essential for determining the magnetospheric activity. Later Kane and Echer (2007) confirmed the importance of Bz component. Our result is in agreement with the results by Kane and Echer (2007) in the sense that for intense storms the larger negative Bz gives the stronger negative Dst and the solar wind velocity possibly does not play a significant geoeffective role. Recently Mansilla, (2008) and Echer et al., (2008) confirmed that peak Dst is correlated to the maximum negative Bz component of the IMF better than the maxima of solar wind number density D and solar wind speed V. Previous results on the correlation between Dst and V were also similar (Crooker and Gringauz, 1993, Papitashvili et al 2000). In our case we have investigated correlation of GI (Kp, ap and AE) in addition to Dst correlations and found that only a moderate correlation between Dst and V exists (Fig. 4h). However Kp, ap and AE indices show a good correlation with V (Figs. 5h, 6h and 7h).

A number of studies for the properties and their correlations of MCs with plasma and field parameters have been performed in the past (Gonzalez et al. 1998; Echer et al., 2005; Gopalswamy et al., 2008a). Gonzalez et al., (1998) found a good correlation between the peak magnetic field and the peak speed in 30 MCs (correlation coefficient = 0.75). However, they did not find a good correlation between the speed and field strength for non cloud ejecta. Echer et al., (2005) confirmed this results but the correlation coefficient was smaller (correlation coefficient = 0.35) for a set of 149 MCs. Gopalswamy et al. (2008a) reported an intermediate correlation (correlation coefficient = 0.56) for a set of 99 MCs in cycle23.

Gonzalez and Tsurutani (1987) used ISEE-3 field and plasma data to determine an empirical relation which states that a necessary interplanatry condition for an intense GMS is the presence of an intense southward component of IMF Bz > -10 nT and the interplanetary dusk-ward electric fields greater than 5 mV/m over a period exceeding 3 hours. Although this empirical relationship was determined for a limited data interval during solar maxima, it appears to hold during solar minimum as well (Tsurutani and Gonzalez, 1995). In our study of 91 intense GMSs we have calculated the average values of these parameters and found a good agreement during rising, maximum and decay phase of solar cycle 23 (Table 2). Along with Bz, VBz is also an appropriate variable relevant for Dst changes (Wu and Lepping, 2002; Srivastava and Venkatakrishnan, 2004; Kane, 2005; Singh et al., 2006) which is confirmed in the present investigation for 91 intense GMSs during solar cycle 23 (Table 2).

There have been only few attempts to relate statistically the various geomagnetic indices (Davis and Pathasarathy, 1967; Campbell, 1979; Saba et al., 1997). Through these studies, useful information has emerged for a better understanding of the storm/substorm relationship. Davis and Pathasarathy (1967) found that the peak Dst index values correlate the best with the time integral of AE during the preceding 10 hours from peak Dst. In our investigation a moderate correlation (R = 0.554) has been obtained between peak Dst and AE-index. Saba et al., (1997) found that the annual average of AE observed in 1974 (near solar minimum) is greater than in 1979 (solar maximum), whereas average Dst is greater in 1979, a year characterized by intense solar transients. The higher occurrence of substorms is responsible for a higher AE, whereas the occurrence of several storms gives a higher Dst for 1979. The correlation coefficient of ap-AE indices is in general the highest, as the magnetometers that monitor both indices are close, and is surpassed only by the ap-Dst correlation during GMSs, when the influence of the ring current is dominant. In our study a high positive correlation between peak values of ap and AE indices is obtained (see Fig. 3(e)) whereas a high negative correlation between ap and Dst indices is obtained during cycle 23 (see Fig. 3(b)).

GMSs are interesting phenomena that result as a final element of a chain of processes starting on the Sun affects the solar wind and interplanetary medium and then reaches the Earth. The ability to predict the occurrence of GMSs on the basis of solar and interplanetary space observations is the basic requirement of investigations. A simple logistic regression model was implemented (Srivastava, 2005a, 2005b) for predicting the occurrence of intense/super-intense GMSs based on a number of solar and interplanetary variables. The results indicate that the model can be used for predicting the occurrence of intense GMSs although it is only moderately successful in predicting super-intense storms. Singh et al., (2006) presented a regression analysis of peak values of Dst of nine intense GMSs versus the peak values of solar wind plasma/field parameters and their various functions during solar cycle 23 and tried to find out interpretation which may be used to obtain the magnitude of storms. In our study we have tried to find out some linear regression equations (see Figs. 4 to 7) for space weather predictions using peak values of GI and various plasma and field parameters of 91 intense GMSs observed during solar cycle 23.

Gonzalez and Echer, (2005) determined that the average delay between the peak Bz component of IMF and the peak Dst index values is ≈ 2 h for 64 GMSs during the period 1997-2002. Recently Gupta and Badruddin (2009) presented a lag/lead time analysis and found a time lag of 1–3 hr between the amplitude of Bz component of IMF and Dst index. In our study the average value comes out ≈ - 4 hr for 91 intense GMSs during the rising, the maximum, the decay and total solar cycle 23. It can be seen the average values of T, V and β shows a lead time with minimum Dst while other plasma and field parameters shows leg time during solar cycle 23 and its phases (Table 4).

We have also tried to find out the solar sources of three GMSs from three different phases. All the GMSs are associated with large flares and halo CMEs and our identified solar sources are in agreement with the results obtained in the past (Wang et al., 2002; Zhang, et al., 2007). Srivastava and Venkatakrishnan, (2004) investigated that fast full-halo CMEs associated with strong flares originating from a favorable location (i.e., close to the central meridian and low and middle latitudes) are the most potential candidates for producing strong ram pressure at the Earth's magnetosphere and hence intense GMSs. Gopalswamy et al., (2007) also studied the halo CMEs and found that disk halos are likely to arrive at the Earth and cause GMSs, while limb halos only impact the Earth with their flanks and hence are less geoeffective. In our case we have found out that all the three GMSs from the rise, the maximum and the decay phases are associated with halo CMEs which is in agreement with the statement above (Table 5 and Figs. 12, 13 and 14).

In order to explain GMSs in terms of enhanced solar wind magnetospheric coupling activity various researchers have come out with a number of suggestions. Axford and Hines, (1961) suggested that there is a viscous interaction between the solar wind and the Earth's magnetopause resulting in an energy transfer. Oppositely directed magnetic field lines fuse into each other which convert the magnetic energy into heat which warms and accelerates the plasma. When magnetic topology changes, totally unconnected regions may exchange plasma, mass, momentum and energy as explained by Hughes (1995). In the case of Earth's magnetosphere, the magnetic reconnection is more effective when the Bz component of IMF is south directed (Dungey, 1961). For intense magnetic storms which are the outcome of the geoeffective impact of solar wind the solar wind speed and the IMF intensity must be substantially higher and the field must be south directed for a considerable length of time (Gonzales et al., 1999).

Intense and super-intense GMSs produce disturbances in the ionosphere-thermosphere system and can cause communication failure and navigational errors. Heating and subsequent expansion of the thermosphere during such storms could produce extra drag on the low earth orbiting satellites and can reduce their lifetimes significantly. Super-intense GMSs like the one reported on September 1-2, 1859 if were to occur today would adversely affect the space-weather conditions and catastrophic devastation (Tsurutani et al. 2003).

**ACKNOWLEDGEMENT**

We are grateful to various catalogues available in publications and on the internet, especially WDC, CDAW and NSSDC/NASA's OMNI, SOHO/MDI, SOHO/EIT, SOHO/LASCO. Two of the authors (NCJ and NSB) wish to thank UGC, New Delhi for financial assistance under RFSMS (Research Fellowship in Science for meritorious students) scheme.

--------

## CAPTION OF FIGURES

**Figure 1(a).** Variation of intense CMEs, Hα flares, SXR flares, SNs and GMSs (Dst ≤ -100 nT) during solar cycle 23 (1996-2008).

**Figure 1(b).** Pie chart diagram showing the GMSs during rising, maximum and decay phase of solar cycle 23.

**Figure 2.** Relation between yearly values of various solar activity features with yearly numbers of GMSs during solar cycle 23 (1996-2008): (a) GMSs vs sunspot numbers (SNs) (b) GMSs vs SXR flares (c) GMSs vs Hα flares (d) GMSs vs coronal mass ejections (CMEs). The correlation coefficients (R), equation of regression line and straight line fit (dash line) are represented in each panel.

**Figure 3.** Relation between peak values of various GI for 91 GMSs during solar cycle 23 (1996-2008): (a) Kp vs Dst (b) ap vs Dst (c) AE vs Dst (d) AE vs Kp (e) AE vs ap. The correlation coefficients (R) and straight line fit (broken lines) are represented on each panel.

**Figure 4.** Relation between peak Dst index and peak values of various plasma and field parameters: (a) Dst vs FMA ($|B|$) (b) Dst vs Bt (c) Dst vs By (d) Dst vs Bz (e) Dst vs $\sigma_B$ (f) Dst vs T (g) Dst vs D (h) Dst vs V (i) Dst vs P (j) Dst vs E (k) Dst vs β (l) Dst vs VBz. The correlation coefficients (R), equation of regression line and straight line fit (broken lines) are represented in each panel.

**Figure 5.** Relation between peak ap index and peak values of various plasma and field parameters: (a) ap vs FMA ($|B|$) (b) ap vs Bt (c) ap vs By (d) ap vs Bz (e) ap vs $\sigma_B$ (f) ap vs T (g) ap vs D (h) ap vs V (i) ap vs P (j) ap vs E (k) ap vs β (l) ap vs VBz. The correlation coefficients (R), equation of regression line and straight line fit (broken lines) are represented in each panel.

**Figure 6.** Relation between peak Kp index and peak value of various plasma and field parameters: (a) Kp vs FMA ($|B|$) (b) Kp vs Bt (c) Kp vs By (d) Kp vs Bz (e) Kp vs $\sigma_B$ (f) Kp vs T (g) Kp vs D (h) Kp vs V (i) Kp vs P (j) Kp vs E (k) Kp vs β (l) Kp vs VBz. The correlation coefficients (R), equation of regression line and straight line fit (broken lines) are represented in each panel.

**Figure 7.** Relation between peak AE index and peak value of various plasma and field parameters: (a) AE vs FMA ($|B|$) (b) AE vs Bt (c) AE vs By (d) AE vs Bz (e) AE vs $\sigma_B$ (f) AE vs T (g) AE vs D (h) AE vs V (i) AE vs P (j) AE vs E (k) AE vs β (l) AE vs VBz. The correlation coefficients (R), equation of regression line and straight line fit (broken lines) are represented in each panel.

**Figure 8.** Distribution of the GI: (a) Dst index (b) Kp index (c) ap index (d) AE index. The average and median of the distributions are represented and also marked with solid and broken lines respectively. Standard deviation (SD) has also been represented.

**Figure 9.** Distribution of the plasma and field parameters: (a) FMA ($|B|$) (b) Bt (c) By (d) Bz (e) $\sigma_B$ (f) T (g) D (h) V (i) P (j) E (k) β (l) VBz. The average and median values of the distribution are marked with solid and broken lines respectively. Standard deviation (SD) has also been represented.

**Figure 10**. Distribution of time difference between the peak time of geomagnetic disturbance (i.e. minimum Dst index) and other GI. (a) Kp and ap indices (b) AE index. Average, median and mode values of the distribution are marked with solid and broken lines respectively. Mode and Standard deviation (SD) have also been represented.

**Figure** 11. Distribution of time difference between the peak time of geomagnetic disturbance (i.e. minimum Dst index) and various plasma and field parameters (a) FMA ($|B|$) (b) Bt (c) By (d) Bz (e) $\sigma_B$ (f) T (g) D (h) V (i) P (j) E (k) β. The average, median and mode values of the distribution are marked with solid and broken lines respectively. Standard deviation (SD) has also been represented.

**Figure 12. On the left panel, from top to bottom**: SOHO MDI sunspot image (November 07, 1997; 04:47 UT); EIT 195 snapshot taken on November 07, 1997 at 05:58 UT showing the flare; LASCO C2 and LASCO C3 snapshots taken on the same day at 06:10 UT and 07:46 UT showing the full halo CME. **On the right panel from top to bottom**: IMF parameters ($\sigma_B$, Bz, By, Bt and $|B|$), GI (ap, AE, Dst, and Kp indices) Bx, Plasma parameters (β, E, P, V, D and T) for the period of November 6 to November 10, 1997.

**Figure 13. On the left panel, from top to bottom**: SOHO MDI sunspot image (October 19, 2001; 16:03 UT); EIT 195 snapshot taken on October 19, 2001, at 16:36 UT showing the flare; LASCO C2 and LASCO C3 snapshots taken on the same day at 17:26 UT and 18:05 UT showing the full halo CME. **On the right panel, from top to bottom**: IMF parameters ($\sigma_B$, Bz, By, Bt and $|B|$), GI (ap, AE, Dst, and Kp indices) Bx, Plasma parameters (β, E, P, V, D and T) for the period of October 20 to October 24, 2001.

**Figure 14. On the left panel, from top to bottom**: SOHO MDI sunspot image (December 13, 2006; 01:39 UT); EIT 195 snapshot taken on 13 December, 2006, at 02:36 UT showing the flare; LASCO C2 and LASCO C3 snapshots taken on the same day at 02:54 UT and 03:42 UT showing the full halo CME. **On the right panel, from top to bottom**: IMF parameters ($\sigma_B$, Bz, By, Bt and $|B|$), GI (ap, AE, Dst, and Kp indices) Bx, Plasma parameters (β, E, P, V, D and T) for the period of December 13 to December 18, 2006.